\documentclass[fleqn,twoside]{article}
\usepackage{espcrc2}

\usepackage{graphicx}

\newcommand{\AmS}{{\protect\the\textfont2
  A\kern-.1667em\lower.5ex\hbox{M}\kern-.125emS}}

\title{Status of the BAIKAL neutrino project}

\author{
V.Balkanov\address[INR]{Institute for Nuclear Research, Moscow, Russia},
I.Belolaptikov\address[JINR]{Joint Institute for Nuclear Research, Dubna, Russia},
N.Budnev\address[IGU]{Irkutsk State University, Irkutsk,Russia},
L.Bezrukov\addressmark[INR],
A.Chensky\addressmark[IGU],
D.Chernov\address[MSU]{Skobeltsyn Institute of Nuclear Physics  MSU, Moscow, Russia},
I.Danilchenko\addressmark[INR],
Zh.-A.Dzhilkibaev\addressmark[INR],
G.Domogatsky\addressmark[INR],
A.N.Dyachok\addressmark[IGU],
O.Gaponenko\addressmark[INR],
O.Gress\addressmark[IGU],
T.Gress\addressmark[IGU],
A.Klabukov\addressmark[INR],
A.Klimov\address[KUR]{Kurchatov Institute, Moscow, Russia},
S.Klimushin\addressmark[INR],
K.Konischev\addressmark[INR],
A.Koshechkin\addressmark[INR],
L.Kuzmichev\addressmark[MSU],
V.Kulepov\address[NOV]{Nizhni Novgorod State Technical University},
Vy.Kuznetzov\addressmark[INR],
B.Lubsandorzhiev\addressmark[INR],
S.Mikheyev\addressmark[INR],
M.Milenin\addressmark[NOV],
R.Mirgazov\addressmark[IGU],
N.Moseiko\addressmark[MSU],
E.Osipova\addressmark[MSU],
A.Pavlov\addressmark[IGU],
G.Pan'kov\addressmark[IGU],
L.Pan'kov\addressmark[IGU],
A.Panfilov\addressmark[INR],
Yu.Parfenov\addressmark[IGU],
E.Pliskovsky\addressmark[JINR],
P.Pokhil\addressmark[INR],
V.Polecshuk\addressmark[INR],
E.Popova\addressmark[MSU],
V.Prosin\addressmark[MSU],
M.Rosanov\address[LEN]{St.Peterburg State Marine University, St.Peterburg, Russia},
V.Rubtzov\addressmark[IGU],
Y.Semeney\addressmark[IGU],
B.Shaibonov\addressmark[INR],
Ch.Spiering\address[DESY]{DESY--Zeuthen, Zeuthen, Germany},
O.Streicher\addressmark[DESY],
B.Tarashanky\addressmark[IGU],
R.Vasiliev\addressmark[JINR],
E.Vyatchin\addressmark[INR],
R.Wischnewski\addressmark[DESY],
I.Yashin\addressmark[MSU],
V.Zhukov\addressmark[INR]
}

\begin{document}

\begin{abstract}
 We review the present status of the Baikal Neutrino Project
and present results on upward going atmospheric neutrinos, 
results of a search for high energy extraterrestrial neutrinos as well as
preliminary results of searching for acoustic signals from
EAS in water. We describe the moderate upgrade of NT-200
planned for the next years and discuss a possible detector
on the Gigaton scale.  

\vspace{1pc}
\end{abstract}

\maketitle

\section{DETECTOR AND SITE}

The Baikal Neutrino Telescope  is operated in Lake 
Baikal, Siberia,  at a depth of \mbox{1.1 km}. 
The present stage of the telescope,
NT-200 \cite{APP}, was put into operation 
at April 6th, 1998 and consists of 192
optical modules (OMs). 
An umbrella-like frame carries  8 strings,
each with 24 pairwise arranged OMs.
Three underwater electrical cables and one optical cable connect the
detector with the shore station. 

The OMs are grouped in pairs along the strings. They contain 
37-cm diameter {\it QUASAR} - photo multipliers (PMs) 
which have been developed
specially for our project \cite{OM2}. The two PMs of a
pair are switched in coincidence in order to suppress background
from bioluminescence and PM noise. A pair defines a {\it channel}. 

A {\it muon-trigger}
is formed by the requirement of \mbox{$\geq N$ {\it hits}}
(with {\it hit} referring to a channel) within \mbox{500 ns}.
$N$ is typically set to 
\mbox{3 or 4.} For  such  events, amplitude and time of all fired
channels are digitized and sent to shore. 
A separate {\em monopole trigger} system searches for clusters of
sequential hits in individual channels which are
characteristic for the passage of slowly moving, bright
objects like GUT monopoles.

 Lake Baikal deep water is characterized by an
absorption length of $L_{abs}$(480 nm)=20 $\div $24 m, a scattering
length of $L_s=$30 $\div $70 m and a strongly anisotropic scattering
function $f(\theta)$ with a mean cosine of the scattering angle 
$\overline{\cos}(\theta)=0.85 \div 0.9$.
\begin{figure}[htb]
\vspace{-0.7cm}
\includegraphics[width=8.3cm,height=7.cm]{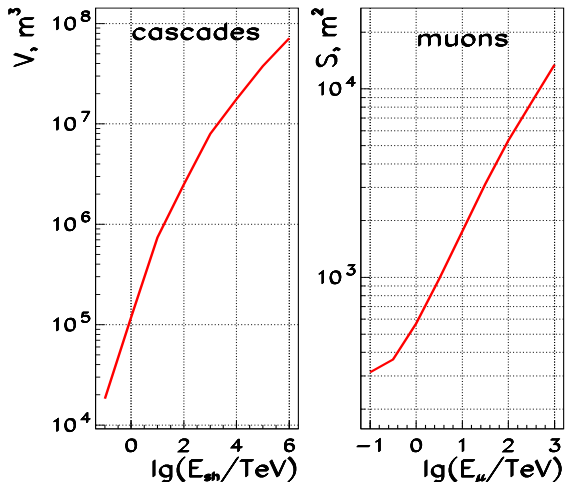}
\vspace{-0.7cm}
\caption{
Detection volume (left) and detection area (right) of 
a single BAIKAL OM for neutrino induced
high energy cascades and high energy muons,
respectively. 
}
\end{figure}
Fig.1 shows the cascade detection volume and
the muon detection area of a single BAIKAL
OM. Here, we define detection area and detection
volume by the condition that the mean number
of photoelectrons has to be $\ge$1. In contrast
to underground detectors, open configurations
in highly transparent media like water or ice allow
to observe a huge volume beyond their geometrical
boundaries. 
The detection volume of an OM rises
from 1$\cdot$10$^5$ m$^3$ for 1 TeV to
 7$\cdot$10$^7$ m$^3$ for 1 EeV
cascade energy.

Here we present results on upward going atmospheric
neutrinos, a search for
high energy extraterrestrial neutrinos, and
preliminary results of a search for acoustic signals from
EAS in water.  Also an upgrade of NT-200 towards the Gigaton scale
detector is dicussed.

\section{A SEARCH FOR UPWARD GOING ATMOSPHERIC NEUTRINOS}

The signature of neutrino induced events is a muon crossing the
detector from below. The reconstruction algorithm is based on the
assumption that the light radiated by the muons is emitted under
the Cherenkov angle with respect to the muon path. 
We don't take into account light scattering because the characteristic
distances for muons induced by atmospheric neutrinos which
trigger the detector
do not exceed 1$\div$2 scattering lengths of light in Baikal water 
(see above).

The algorithm uses a single muon model to reconstruct
events. 
We apply procedures to reject hits, which are very likely
due to dark current or water luminosity as well as hits
which are due to  
showers and have large time delays with respect to expected
hit times from the single muon Cherenkov light.

Determination of the muon trajectory is based on the minimization of a
$\chi^2$ function with respect to measured
and calculated times of hit channels.
As a result of the $\chi^2$ minimization we obtain the track parameters
($\theta$, $\phi$ and spatial coordinates). 
\begin{figure}[htb]
\vspace{-2mm}
\includegraphics[width=7.5cm]{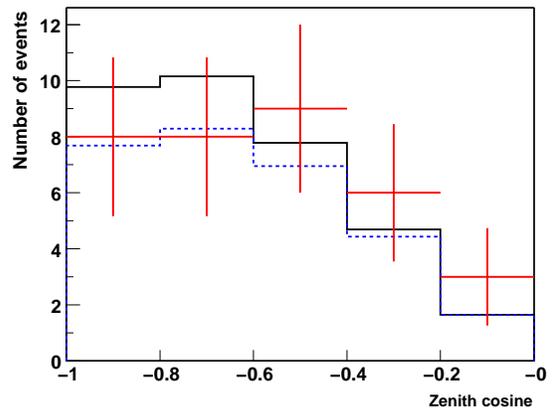}
\vspace{-1.2cm}
\caption{
Experimental angular distribution of events 
reconstructed and accepted as upward going muons
 (solid crosses). The MC expectation for atmospheric neutrinos
is shown for the case of no oscillation (solid line) and for
oscillations with maximal mixing and $\Delta m^2 =$ 3$\cdot$
10$^{-3} \,$ eV$^2$ (dashed line).
(MC distribution do not include the background due to atmospheric 
muons.)}
\vspace{-8mm}
\label{fig:angul}
\end{figure}

Since the algorithm yields a large fraction of events which are
reconstructed as upward going muons, 
wrongly reconstructed tracks have been removed by a series 
of quality cuts \cite{APP2,SUDB00}.

The efficiency of reconstruction and ``fake'' event
rejection and the correctness of the MC background estimation have been
tested with a sample of 1.1$\cdot$10$^8$ MC-generated events from
atmospheric muons (that is twice the whole experimental sample,
see below) and with MC-generated events due to atmospheric neutrinos.
Muons passage through rock and water has been generated by
transport code MUM \cite{SOK}.

For the MC background sample we found 3 surviving events reconstructed
as upward going muons and passing the quality cuts. 
The analysis of these events shows that they are  
complex and contain one or two muons with showers 
(shower energy $\ge$ 100 GeV). 
These events are reconstructed 
close to horizon ($\cos (\theta)>-$0.3).

Data taken with NT-200 between  April 1998 and February 1999  
cover 234 days life time. 
From the total of 1.67$\cdot$10$^8$ triggers 
($N_{\mbox{\rm hit}}\ge$4)
recorded in 234 days, 5.3$\cdot$10$^7$ events with 
$N_{\mbox{\rm hit}}\ge$6
at $\ge$ 3 strings have been selected for this analysis. 
These events have been reduced to 34 upward tracks.
The MC estimations of the
number of expected upward going events due to neutrino induced muons 
give
34 and 29.4 events in case of absence of neutrino oscillations and
including oscillations with maximal mixing angle and 
$\Delta m^2 = $3$\cdot$10$^{-3}$ eV$^2$, respectively.
Angular distributions of both experimental and MC samples are presented in
Fig.\ref{fig:angul}.

\begin{figure}[h]
\vspace{-6mm}
\includegraphics[width=7.5cm]{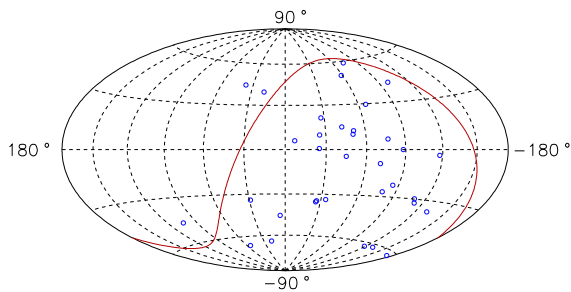}
\vspace{-1.2cm}
\caption{
The celestial distribution of the experimental neutrino sample in galactic
coordinates. The thick solid line indicates the horizon in 
equatorial coordinates.}
\vspace{-8mm}
\label{fig:celes}
\end{figure}

Fig.\ref{fig:celes} shows the celestial distribution of selected
upward going muons with NT-200 in galactic coordinates.

\section{A SEARCH FOR EXTRATERRESTRIAL 
HIGH ENERGY NEUTRINOS}

The used search strategy for high energy neutrinos relies
on the detection of the Cherenkov light emitted by the electro-magnetic 
and (or) hadronic particle cascades and high energy muons
produced at the neutrino interaction
vertex in a large volume around the neutrino telescope \cite{VEN01,APP3}.
A cut is applied which accepts only time patterns 
corresponding to upward traveling light signals. 

For this analysis we used 
1.56$\cdot$10$^5$ events with $N_{\mbox{\rm hit}}>$10. 

\begin{figure}[htbp]
\vspace{-0.6cm}
 \includegraphics[width=7.5cm]{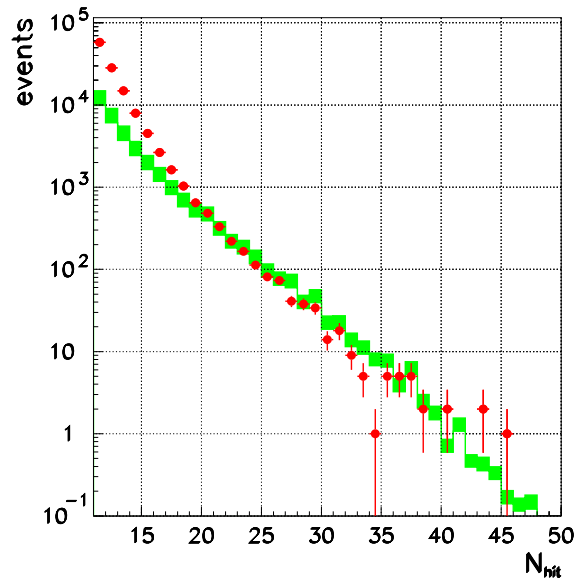}
\vspace{-0.7cm}
\caption{Distribution of hit channel multiplicity; dots - experiment,
hatched boxes - expectation from brems and hadronic showers
produced by atmospheric muons.}
\end{figure}

Fig.4 shows the $N_{\mbox{\rm hit}}$ distribution for experiment (dots)
as well as the one expected for the background from brems- and
hadronic high energy showers produced by atmospheric muons (boxes).
The experimental distribution is consistent with the 
background expectation for $N_{\mbox{\rm hit}}>$18. 
The discrepancy at lower $N_{\mbox{\rm hit}}$ values is caused
by the contribution of atmospheric muons close to horizon
as well as by low energy showers from $e^+e^-$ pair production.
These effects have not been included in the present MC sample,
but have been taken into account in separate simulations which
are consistent with experimental data.
No statistically significant excess over background
expectation from atmospheric
muon induced showers has been observed.
Since no events with $N_{\mbox{\rm hit}}>$45 
are found in our data, we can derive upper limits on the flux of 
high energy neutrinos which would produce events with $N_{\mbox{\rm hit}}>$50. 

The detection volume $V_{eff}$ for neutrino produced events 
with $N_{\mbox{\rm hit}}>$50 which fulfill all trigger 
conditions was calculated as a function of neutrino energy
and zenith angle $\theta$.  $V_{eff}$ rises 
from 2$\cdot$10$^5$ m$^3$ for 10 TeV up to 6$\cdot$10$^6$ m$^3$    
for $10^4$ TeV and significantly exceeds the geometrical volume
$V_{g} \approx$ 10$^5$ m$^3$ of NT-200. 

Given an $E^{-2}$ behaviour of the neutrino spectrum and a
flavor ratio 
$(\nu_e+\tilde{\nu_e}):(\nu_{\mu}+\tilde{\nu_{\mu}})=1:2$,  
the combined 90\% C.L. upper limit obtained with the Baikal 
neutrino telescopes 
NT-200 (234 days) and NT-96 \cite{APP3} (70 days) 
is:

\begin{equation}
\Phi_{(\nu_e+\tilde{\nu_e})}E^2<(1.3 \div 1.9)\cdot10^{-6} 
\mbox{cm}^{-2}\mbox{s}^{-1}\mbox{sr}^{-1}\mbox{GeV}
\end{equation}
where the upper value allows for the highest light scattering
observed over many seasons.

\begin{figure}[htb]
\vspace{-0.7cm}
\includegraphics[width=8.3cm]{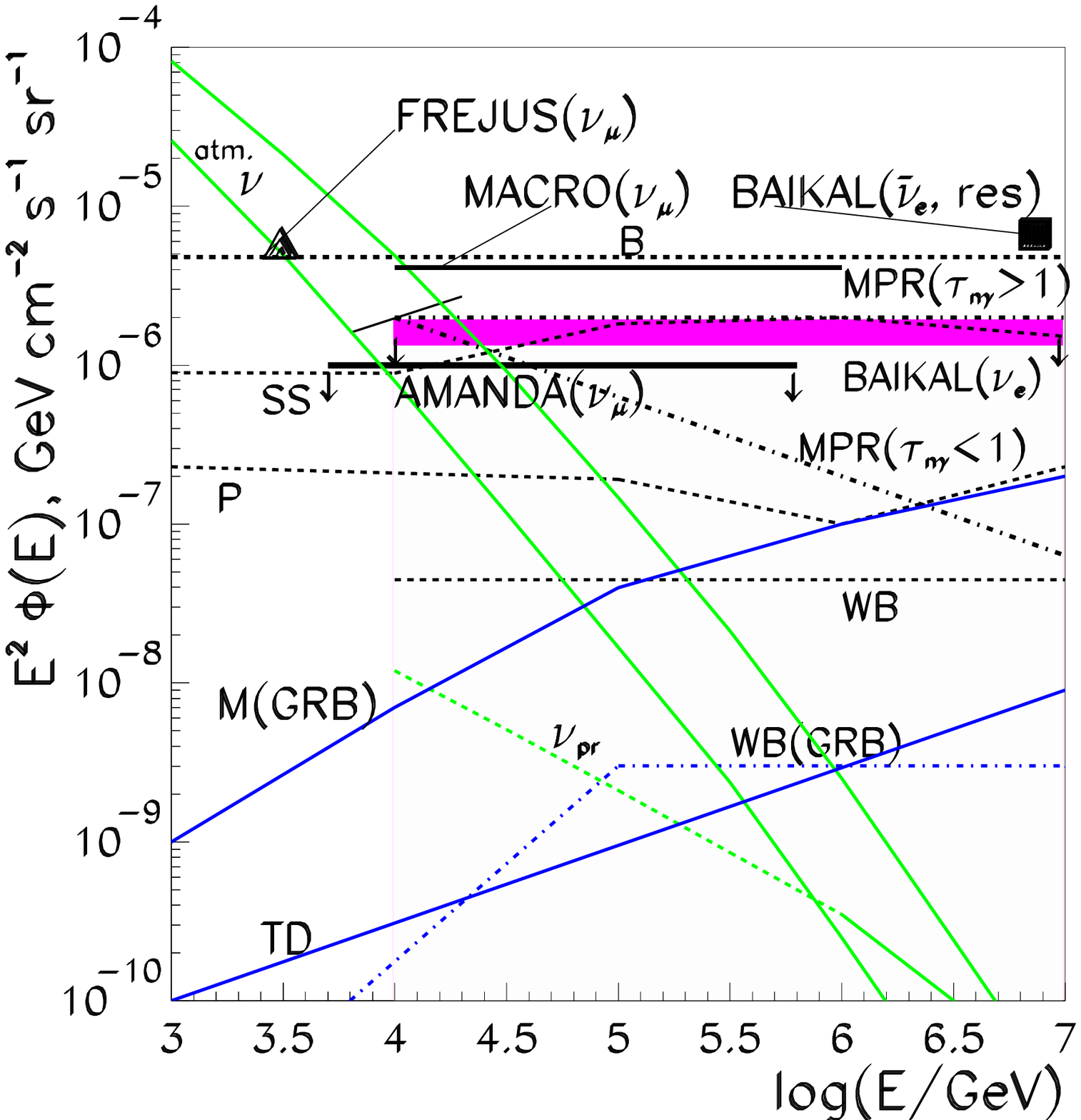}
\vspace{-0.7cm}
\caption{Experimental upper limits on the neutrino fluxes as well as flux 
predictions in different models of neutrino sources (see text).
}
\end{figure}

Fig.5 shows the upper limits on the isotropic diffuse neutrino
flux obtained by $\,$ BAIKAL $\,$(this work), 
\mbox{AMANDA} \cite{AMANDA2}, MACRO \cite{MACRO}
and FREJUS \cite{FREJUS} (triangle) 
as well as the atmospheric conventional neutrino \mbox{fluxes \cite{VOL}} 
from horizontal and vertical directions (upper and lower curves,
respectively) and the atmospheric prompt neutrino flux  
\cite{PROMT} (curve labeled $\nu_{pr}$). 
Also shown is the model-independent upper limit on the diffuse high energy neutrino flux
obtained by Berezinsky \cite{Ber3} (curve labeled 'B'),
and predictions for diffuse neutrino fluxes
from Stecker and Salamon  \cite{SS} 
('SS') $\,$ and 
$\,$ Protheroe \cite{P} ('P').
Curves labeled 'MPR' and 'WB' show the upper bounds obtained by 
Mannheim et al.
\cite{P98} as well as the upper bound obtained
by Waxman and Bahcall \cite{WB1}, respectively.
Curves labeled 'M(GRB)' and 'WB(GRB)' present the upper bounds for diffuse
neutrino flux from GRBs derived by Mannheim \cite{MANNHEIM} and
Waxman and Bahcall \cite{WB2}. 
The curve labeled 'TD' shows the prediction for neutrino flux from topological
defects due to specific top-down scenario BHS1 \cite{BHS1}.

Our combined 90\% C.L. limit at the W - resonance energy is:
\begin{equation}
\frac{d\Phi_{\bar{\nu}}}{dE_{\bar{\nu}}} \leq (1.4 \div 1.9) \times 
10^{-19} 
\mbox{cm}^{-2}\mbox{s}^{-1}\mbox{sr}^{-1}\mbox{GeV}^{-1}
\end{equation}
and is given by the rectangle in Fig.5.

\section{SEARCH FOR ACOUSTIC SIGNAL FROM EAS CORE IN WATER}

In 2002 the Baikal collaboration has reported  the observation of 
characteristic bipolar acoustic signal in time coincidence with an EAS detected by 
a scintillator array \cite{SUDB00}. The scintillator array was placed on ice cover 
and the hydrophone 
in water, at the depth of 5 m and at a distance of 90 m from the array center. 
In the two years since then experiments with acoustic hydrophones triggered by 
the EAS array have been  continued.

In both experiments the EAS array consisted of seven scintillator counters. Six counters 
formed a circle with radius 80 m in 2001 (50 m in 2002). The seventh counter was 
placed in the center of the circle. Measurements of the number of relativistic particles in 
each counter and the relative time delays allow reconstructing the shower parameters: 
arrival direction, core position and total number of electrons. 

In the experiment of March/April 2001, four hydrophones were located at a distance of 
34 m from the center of the EAS array and formed a square. Ten runs of measurements 
have been carried out. Hydrophones were placed in water at a depth of 4 m in seven runs, 
while they were located in air, 
on the ice cover,
in three runs. 

In the second experiment (March/April 2002), a more compact acoustic antenna was used. 
Three hydrophones were placed 
at the
corners of an equilateral triangle at a distance of 1 m 
each from other. All three hydrophones were in water at a depth of 4 m. The fourth 
hydrophone was located below the triangle center at a depth of 8 m.

On top of our own anntenas, hydrophones from an ITEP/Moscow group have been
operated in both experiments. 
ITEP's 
antennas had four hydrophones of different type, placed on a vertical string and 
separated  from each other by 5 m \cite{ITEP}.

 In order to record an acoustic signal, a multi channel digitizer was used. The EAS 
array triggered the
data acquisition system 
and started the digitizers. The digitizers recorded the 
signal from each hydrophone during 0.1 sec with 
a step width of 
2 $\mu$sec. So, for each shower 
there are four time series 
consisting
of 50 000 measurements of the amplitude of the acoustic 
signal.

The analysis of the time series shows that apart from Gaussian noise, which can be 
parameterized 
by 
average
amplitude and standard deviation $\sigma$, there are large fluctuations 
lasting from tens up to hundreds microseconds. These "perturbations" could be due to real 
acoustic signals produced by some source. 
Selecting "perturbations" with a duration 
between 20 and 100 $\mu$sec 
and an average amplitude 
more than 2$\sigma$, we have found in average 50 "perturbations" in each channel per 
one event. Probably, they are mainly due to ice cracking. 

Combining "perturbations" recorded by the four hydrophones and using the known sound 
velocity 
in Baikal water, 1405 m/sec, one can reconstruct the position of the source in space 
and time. We 
have found 46 events within a time window +/-0.5 msec relative to the EAS trigger, when 
the hydrophones were placed in water, and 13 events in the case of hydrophone 
location in air. 

The expected background was estimated assuming that the sound sources are 
independent and 
uniformly distributed in ice. Simple Monte Carlo calculations predict 26 and 12 events 
respectively. One can say there is excess of events within time window +/- 0.5 msec at the 
level of 2$\sigma$ in the case of hydrophones located in water. The analysis 
of the 2002 experiment 
is in progress.

\section{NT-200$+$ AND BEYOND}

Recently derived 
upper limits on $\nu_{\mu}$ (AMANDA) and $\nu_{e}$ (BAIKAL) fluxes are
about $E^2 F(\nu) \approx 10^{-6}$ cm$^{-2}$ s$^{-1}$ sr$^{-1}$ GeV
and  belong to the region of most optimistic theoretical predictions.
A flux sensitivity at the
level of  $E^2 F(\nu) \leq 10^{-7}$ cm$^{-2}$ s$^{-1}$ sr$^{-1}$ GeV
which would test a variety of other models,
requires  detection volumes of a few dozen Mtons.

We envisage an upgrade of NT-200 to the scale of a few dozen Mton by
three sparsely instrumented distant outer strings. The basic principle
will be the search for cascades produced in a large volume below
NT-200. This configuration, christened NT-200$+$, will not only
result in an increased detection volume for cascades, but also allow
for a precise reconstruction of cascade vertex and energy within
the volume spanned by the outer strings.

\begin{figure}[h]
\vspace*{-2.0mm} 
\includegraphics[width=7.5cm,height=9cm]
{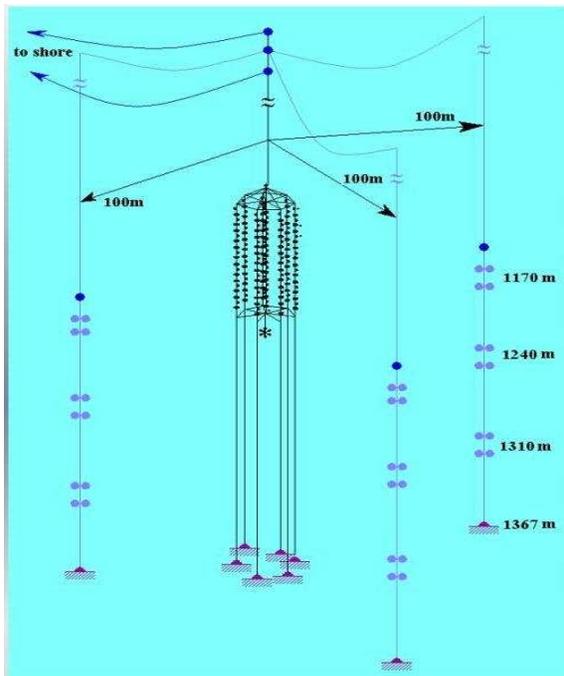} 
\label{fig:detector_layout}
\caption{Sketch of NT-200+.}
\end{figure}
A schematic view of NT-200+ is shown in Fig.6.
It will comprise the neutrino telescope NT-200 itself as well as three
140 m long outer strings with 3 pairs of OMs spaced vertically by 70 m, 
and two pairs of OMs spaced by 6 m and arranged at one of the eight strings of 
NT-200, 140 m below it.
The outer strings are arranged 
at a distance of 100 m around NT-200 at the edges of an equilateral
triangle.
Their top OMs are located
at the level of the bottom OMs of NT-200.
A water volume of $4.4 \cdot 10^6$ m$^3$ 
is surrounded by the outer strings and NT-200. 
\begin{figure}[htb]
\vspace*{-2.0mm} 
\includegraphics[width=7.5cm,height=8cm]
{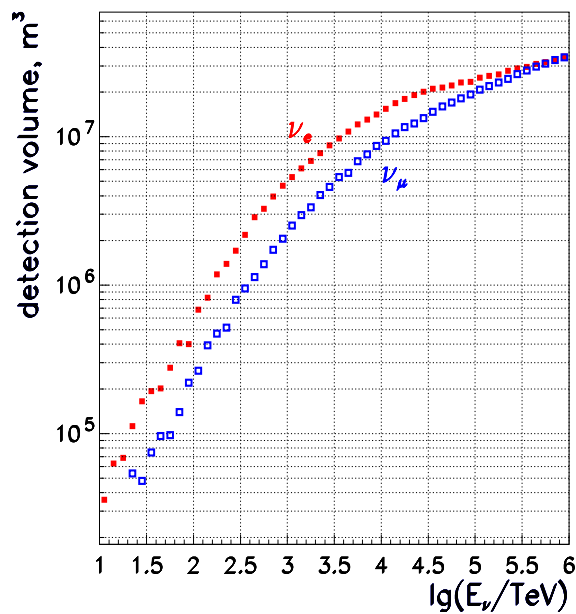} 
\label{fig:detector_layout}
\caption{Detection volume of NT-200$+$ for $\nu_e$ and $\nu_{\mu}$ 
events which survive all cuts.}
\end{figure}

Moderate event selection requirements allow to achieve a large detection
volume for neutrino events and to suppress background effectively.
The detection volumes for isotropic $\nu_e$ and $\nu_{\mu}$ fluxes 
are shown in Fig.7. 
The value of $V_{eff}$ for $\nu_e$ induced events rises 
from 7$\cdot$10$^5$ m$^3$ for 100 TeV up to 4$\cdot$10$^7$ m$^3$    
for $10^6$ TeV. 
Normalized energy distributions of  expected event rates  from 
$\nu_e$ and $\nu_{\mu}$ fluxes following an inverse power law
with  spectral index $\gamma=2$ are presented in Fig.8.
Most of the expected events would be produced by  neutrinos
from the energy range $10^2 \div 10^5$ TeV, with a mean energy 
around 1 PeV.
\begin{figure}[htb]
\vspace*{-2.0mm} 
\includegraphics[width=7.5cm,height=8cm]
{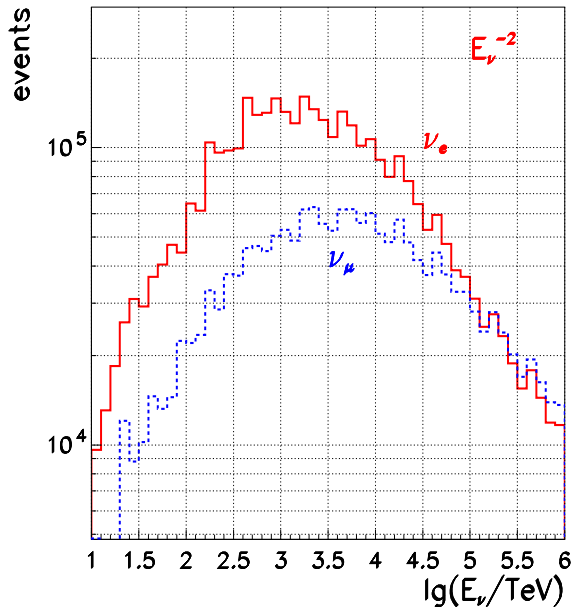} 
\label{fig:detector_layout}
\caption{Energy distribution of expected events induced by
diffuse $\nu_e$ and $\nu_{\mu}$ fluxes.}
\end{figure}

Assuming $\gamma = 2$ and flavor ratio 
$(\nu_e+\tilde{\nu_e}):(\nu_{\mu}+\tilde{\nu_{\mu}})=1:2$  
and a $\nu_e$ flux of
\begin{equation}
\Phi_{(\nu_e+\tilde{\nu_e})}E^2=3.5 \cdot10^{-7} 
\mbox{cm}^{-2}\mbox{s}^{-1}\mbox{sr}^{-1}\mbox{GeV}
\end{equation}
a one year expectation would yield 2.4 events.
\begin{figure}[htb]
\vspace*{-2.0mm} 
\includegraphics[width=7.5cm,height=8cm]
{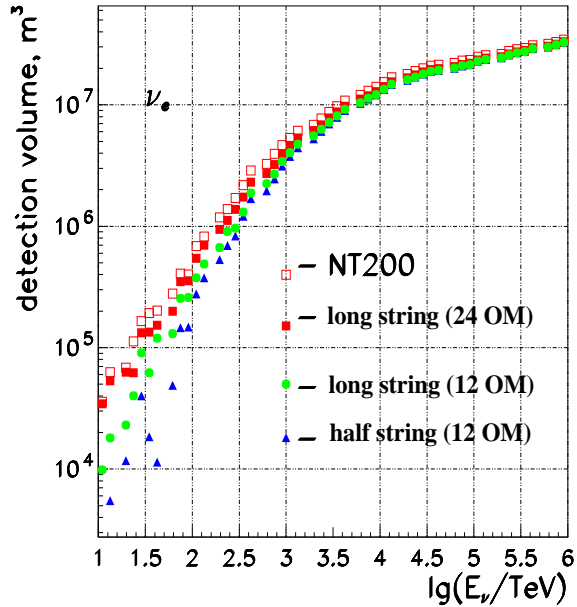} 
\label{fig:detector_layout}
\caption{Detection volumes of different configurations.}
\end{figure}

MC simulations show that the detection volume of NT-200$+$
for PeV cascades varies only moderately, if NT-200 as the
central part of NT-200$+$ is replaced by a single string of OMs.
Fig.9 gives the detection volume for different configurations
as a function of cascade energy. The standard configuration
of NT-200$+$ is marked by empty rectangles. The other
configurations comprises a single string instead of NT-200:
a standard string of 70 m length and 24 OMs (filled rectangles),
a  half string with 12 OMs covering 35 m (dots), and a 70 m
long string sparsely equipped with 12 OMs (triangles).
The configuration with the long  12-OM string shows  
an energy behaviour very close
to the one of NT-200$+$. For neutrino energies  higher than 100 TeV
such a configuration could be used as a basic subarray of a Gigaton
Volume Detector (GVD). Rough estimations show that 0.7 $\div$ 0.9
Gton detection volume for neutrino induced high energy cascades 
may be achieved with about 1300 OMs arranged at 91 strings. A top
view of GVD as well as sketch of one basic subarray are shown in
Fig.10. 
The physical capabilities of GVD at very high energies cover the 
typical spectrum of cubic kilometer arrays. We are presently
working on simulations to optimize the response for TeV muons,
maintaining at the same time the cubic kilometer scale for cascades
with energy above 100 TeV.

%
%
%
%
%
%
%
%
%

\begin{figure}[htb]
\vspace*{-2.0mm} 
\includegraphics[width=8.0cm,height=12.cm]
{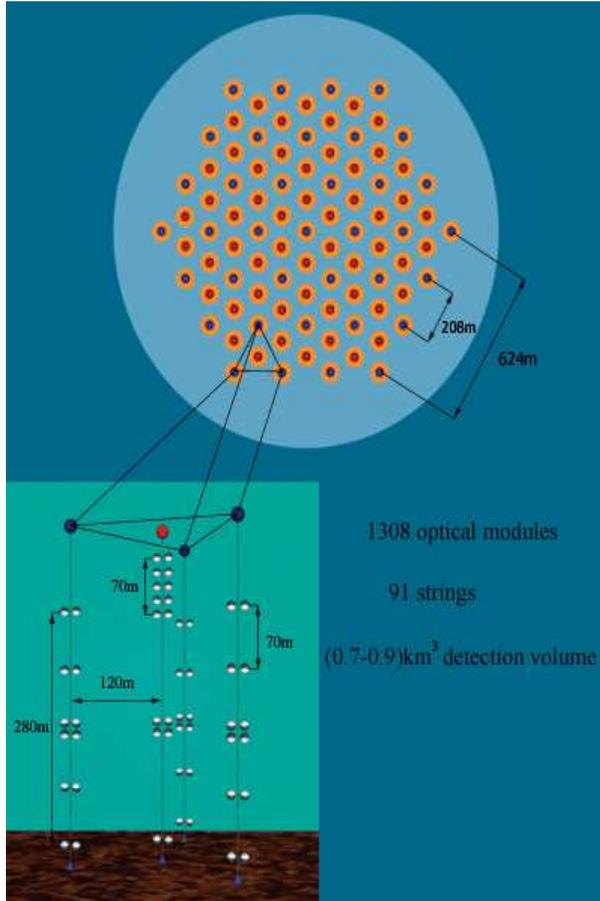} 
\label{fig:detector_layout}
\caption{
Top view of GVD as well as sketch of one of its subarrays.}
\end{figure}

\section{CONCLUSIONS AND OUTLOOK}

The deep underwater neutrino telescope NT-200 in Lake Baikal is 
taking data since April 1998.
Using the first 234 live days, 34 neutrino induced upward going 
muons have been selected. 
The limits on the diffuse high energy ($\nu_e +\bar{\nu}_e$) flux
as well as $\bar{\nu}_e$ flux at the W-resonance energy have been derived.
Also limits on an excess of the muon flux due to WIMP annihilation in the center 
of the Earth and 
on the flux of fast magnetic monopoles have been obtained.
The additional analyzes of the 
experimental data accumulated during 1999-2002 will
allow to decrease these limits by a factor 3-4.

Within the  next few years we plan to upgrade the Baikal neutrino telescope 
to the 10 Mton detector NT-200+ with a sensitivity of approximately 
$3.5\cdot10^{-7}$cm$^{-2}$s$^{-1}$sr$^{-1}$GeV
for a diffuse neutrino flux within the
energy range 10$^2$ TeV $\div$ 10$^5$ TeV.
NT-200+ will search for neutrinos from AGNs, GRBs
and other extraterrestrial sources, neutrinos from cosmic ray
interactions in the Galaxy as well as high energy atmospheric muons
with E$_{\mu}>10$ TeV.

In parallel to this short term goal, we started research \& development
activities towards a Gigaton Volume Detector in Lake Baikal.


\bigskip

{\it This work was supported by the Russian Ministry of Research,the German 
Ministry of Education and Research and the Russian Fund of Fundamental 
Research ( grants } \mbox{\sf 00-15-96794}, \mbox{\sf 02-02-17031}, 
\mbox{\sf 02-02-31005}, \mbox{\sf 02-07-90293}
{\it and} \mbox{\sf 01-02-17227}) 
{\it and by the Russian Federal Program ``Integration'' (project no.} 
\mbox{\sf e0248}).


\begin{thebibliography}{99}
 \vspace{-2mm}

\bibitem{APP} I.A.Belolaptikov {\it et al.}, 
{\it Astropart. Phys.} {\bf 7} (1997) 263. 

\bibitem{OM2} R.I.Bagduev {\it et al.,} {\it Nucl. Instr. Meth.} 
{\bf A420} (1999) 138.

\bibitem{APP2} I.A.Belolaptikov {\it et al.}, 
{\it Astropart. Phys.} {\bf 12} (1999) 75. 

\bibitem{SUDB00} V.A.Balkanov {\it et al.},
{\it Nucl. Phys. Proc. Sup.} {\bf B91} (2001) 438.

\bibitem{SOK} I.A.Sokalski {\it et al.}, 
{\it Phys. Rev.} {\bf D64} (2001) 074015;
hep-ph/0010322. 


\bibitem{VEN01} V.A.Balkanov {\it et al.}, 
{\it Proc. of 9th Int. Workshop on Neutrino Telescopes.
Ed. by Milla Baldo Ceoline. Venezia, March 6-9} (2001) 591;
astro-ph/0105269. 

\bibitem{APP3} V.A.Balkanov {\it et al.}, 
{\it Astropart. Phys.} {\bf 14} (2000) 61. 

\bibitem{AMANDA2} 
E.Andres {\it et al.,} {\it Nucl. Phys Proc. Sup.} {\bf B110}
(2002) 510. 

\bibitem{MACRO}
M.Ambrosio {\it et al.,} {\it Nucl. Phys Proc. Sup.} {\bf B110}
(2002) 519. 

\bibitem{FREJUS} W.Rhode {\it et al.}, {\it Astropart. Phys.} {\bf 4} (1994) 217.

\bibitem{VOL} 
L.Volkova, {\it Yad.Fiz.} {\bf 31} (1980) 1510.

\bibitem{PROMT} 
M.Thunman {\it et al.,} {\it Astr. Phys.} {\bf 5} (1996) 309.

\bibitem{Ber3} V.S.Berezinsky {\it et al.,}
{\it Astrophysics of Cosmic Rays,} North Holland  (1990).

\bibitem{SS} F.W.Stecker and M.H.Salamon, 
astro-ph/9501064 (1995)

\bibitem{P} 
R.Protheroe, {\it The Astron. Soc. of the Pacific}
{\bf 163} (1997) 585; astro-ph/9809144 (1998).

\bibitem{P98} 
K.Mannheim {\it et al.,} astro-ph/9812398 (1998). 

\bibitem{WB1} 
E.Waxman and J.Bahcall, {\it Phys. Rev.} {\bf D59} (1999) 023002.

\bibitem{MANNHEIM} 
K.Mannheim, astro-ph/0010353 (2000). 

\bibitem{WB2} 
E.Waxman and J.Bahcall, {\it Phys. Rev. Lett.} {\bf 78} (1997) 2292.

\bibitem{BHS1} 
P.Bhattacharjee {\it et al.,} {\it Phys. Rev. Lett.} {\bf 69} (1992) 567;
G.Sigl, astro-ph/0008364 (2000). 

\bibitem{ITEP}
V.I.Albul  {\it et al.,} {\it Prib. \& Tech. Eks.} {\bf 3} (2001) 50.




\end{thebibliography}
\end{document}